\documentclass[namedreferences]{solarphysics}
\usepackage[hyperref,optionalrh,solaromanenum]{spr-sola-addons} 
\usepackage{graphicx}                    
\usepackage{amssymb}                     
\usepackage{color}                       

\begin{document}

\begin{article}

\begin{opening}

\title{The Instantaneous Response of Geomagnetic Field, near-Earth IMF and Cosmic-Ray Intensity to Solar Flares}

%
\author[addressref={aff1},corref,email={jouni.j.takalo@oulu.fi; jojuta@gmail.com}]{\inits{J.J.}\fnm{Jouni}~\lnm{Takalo}}

\institute{$^{1}$ Space Physics and Astronomy Research Unit, University of Oulu,
POB 3000, FIN-90014, Oulu, Finland\\
\email{jouni.j.takalo@oulu.fi}}

%
\runningauthor{Jouni J. Takalo}
\runningtitle{Response of GF, IMF and CR to SFI}



\begin{abstract}

We show using superposed epoch analysis (SEA) that the most energetic protons ($>$60 MeV) in near-Earth IMF have a peak almost immediately (less than a day) after peak in solar flare index (SFI), while protons greater than 
10 MeV peak one day after the SFI and protons greater than 1 MeV two days after the SFI.

The geomagnetic indices AU, -AL, PC, Ap and -Dst peak after two to three 
days in SEAs after the peak in SFI. The auroral electrojet indices AU and -AL, however, have only low peaks. Especially, the response of the eastward electrojet, AU, to SFI is negligible compared to other geomagnetic indices.

The SEAs of the SFI and cosmic ray counts (CR) show that the deepest decline in the
CR intensity follows also with 2-3 day lag the maximum of the SFI for the Solar Cycles 20\,--\,24. The depth of the declines are related to the SFI strength of each cycle, i.e, the average decline is about 5\,\% for the Cycles 21 and 22, but only 3\,\% for the Cycle 24. The strongest Cycle 19, however, differs from the other cycles such that it has double-peaked decline and lasts longer than the decline of the other cycles.

The double superposed epoch analyses show that the response of IMF Bv$^2$, which is about two days, and CR to SFI are quite simultaneous, but sometimes Bv$^2$ may peak somewhat earlier than the decline existing in CR.

\end{abstract}

%

\end{opening}

%
\section{Introduction}

The intensity of the cosmic rays has a known period of about 11 years, i.e. the average
length of a solar cycle. Furthermore, the cosmic-ray data experience also
a 22-year or so-called Hale cycle, which is the magnetic polarity cycle of the Sun \citep{Webber_1988, Mavromichalaki_1997, VanAllen_2000, Thomas_2014, Kane_2014, Ross_2019, Takalo_2022_1}.
In addition to these clear solar cycle originated modulations in cosmic rays, Forbush noted a very striking similarity between the variation of the horizontal component of the earth's  magnetic field intensity and of the short-term intensity of the cosmic ray ionization at Cheltenham, Maryland, U.S.A., and Huancayo, Peru, during the very severe magnetic disturbances \citep{Forbush_1937, Hess_1937, Forbush_1938}.
These declines are nowadays called Forbush decreases (FD). According to \cite{Cane_2000}, FD
is usually associated with coronal mass ejections (CME). Earlier, it was believed that solar flares were the cause of this rapid decrease in CR \citep{Lockwood_1971, Venkatesan_1990} but after the discovery of CME, it is realized that instead of the solar flare, CME is the main cause in FDs. \cite{Gopalswamy_2018} observed, however, 
that in most cases, CMEs and solar flares are physically connected to the same active region.

Earlier \cite{Mathpal_2018} used superposed epoch analysis (SEA) (also called Chree analysis \citep{Chree_1913, Chree_1914}) in studies between cosmic ray intensity (CRI) and various solar wind parameters, interplanetary magnetic field (IMF B), geomagnetic storms (GSs), averaged planetary Ap index and sunspot number (SSN) for the
period 2009\,--\,2016 (main part of the Solar Cycle 24) by using their daily mean averages. They found that CRI is anticorrelated with the IMF B. Moreover, using correlation coefficient they found that the delay between the decrease in CRI and increase in IMF B is 0\,--\,2 days such that sometimes decline in CRI is preceding and sometimes succeeding the increase in IMF B. 

\cite{Pokharia_2018} in turn studied IMF Bz-component, solar wind speed (v), product of solar wind speed and Bz (vBz), Kp and Dst indices, and sunspot number (SSN) for the even Solar Cycles 22 and 24 using SEA. In their analysis Dst and Bz were correlated for both Solar Cycles 22 and 24 with correlation coefficient 0.8 . On the other hand, they found that the SSN does not show any relationship with Dst and Kp, while there exists an inverse relation between Dst and the solar wind speed, with some time delay. Furthermore the product vBz was in their analyses more relevant parameter in the production of geomagnetic storms when compared to variables v and Bz separately.

Recently, \cite{Manu_2023} presented a double superposed epoch analysis (DSEA) of the 135 (in SC23) and 61 (in SC24) ICME (interplanetary coronal mass ejection) driven clear geomagnetic storms
activities (DstMin smaller than -50 nT) identified at low, mid and high latitudes using
several geomagnetic indices and the corresponding solar wind variables v, IMF Bz and the
product vBz. Their results revealed that the average reduction in the main IMF driver -vBz (about 24\,\%)
is probably causing nearly the same average storm/activity intensity reductions of geomagnetic field indices at all latitudes (about 25\,\%), though in the AE index the reduction is very small (4\,\%) at high latitudes and in the Kp index the reduction is small (11\,\%) at mid latitudes. They, however, state that the small reductions of AE and Kp are partly due to irregular nature and quasi-logarithmic definition of these indices, respectively.

Here we use SEA in order to study the instantaneous response of geomagnetic field, interplanetary magnetic field (IMF) and cosmic ray intensity (CR) to the rise of the solar flare index (SFI). To this end we use Dst, Ap, PC and auroral electrojet indices AL and AU as geomagnetic field proxies, v and Bz as IMF proxies, and cosmic ray counts mainly from Oulu station, i.e. for the Solar Cycles 20\,--\,24 as a CR proxy. For the analysis of the Solar Cycle 19 we, however, use Climax, Huancayo and Thule stations CR data, because Oulu NM station started operation just in April, 1964. 
This article is organised as follows: Section 2 presents flares and definition of the solar flare index used in this study. In Section 3 we analyse the immense cross-correlation between SFI and geomagnetic indices. Section 4 deals with the superposed epoch analyses and forms the main part of this study. In the Section 5 we give our conclusions.


\section{H$\alpha$ flares and X-ray flares}

The daily solar flare index (SFI), which is directly related to H$\alpha$ flares, used in this study was recently published by \cite{VelascoHerrera_2022}. This database is combined from the records of the Astronomical Institute Ond{\v{r}}ejov Observatory of the Czech Academy of Sciences from 1937\,--\,1976 and the records of the Kandilli Observatory of Istanbul, Turkey from 1977\,--\,2020. It is hard to determine the exact total energy of the solar flare event. That is why the daily and monthly databases are calculated using H$\alpha$ related white-light flares through the formula \citep{Kleczek_1952, Knoska_1984, Ozguc_1989}

\begin{equation}
	Q\,=\,I\times t\:\;     ,
\end{equation}
where {I} is the importance of the flare and t is the duration of the flare in minutes. Furthermore, I consists of two factors, the area of the flare (S, 1, 2, 3, 4) and the brilliance of the flare: F(aint), N(ormal) anf B(right) \citep{Ozguc_2002}. The lengths of the solar cycles used in this study are shown in Table 1.

\begin{table}
\small
\caption{The lengths of the sunspot cycles [fractional years, and year and month] and (starting) sunspot minima for Solar Cycles 19\,--\,24.} 
\begin{tabular}{ c  c  l  c }
  Sunspot cycle    &Fractional    &Year and month     &Cycle length  \\
      number    &year of minimum   & of minimum     &    [years] \\
        \hline   
19    & 1954.3  &1954 April  & 10.5  \\
20    & 1964.8  &1964 October  & 11.7  \\
21    & 1976.5  &1976 June & 10.2  \\
22    & 1986.7  &1986 September  & 10.1  \\
23    & 1996.8  &1996 October  & 12.2  \\
24    & 2009.0  &2008 December  & 11.0  \\ 
25    & 2020    &2019 December 
\end{tabular}
\end{table}

X-ray flares are divided into three different categories; C-, M- and X-flares. These categories follow from to their X-ray brightness in the wavelength range 1\,--\,8 {\AA}ngstr{\"o}ms. Furthermore, each category has nine subdivisions ranging from C1 to C9, M1 to M9, and X1 to X9 in a logarithmic scale. This means that M1 is 10 times stronger than C1 and X1 is 10 times stronger than M1.

\begin{figure} 
 \centerline{\includegraphics[width=1.0\textwidth,clip=]{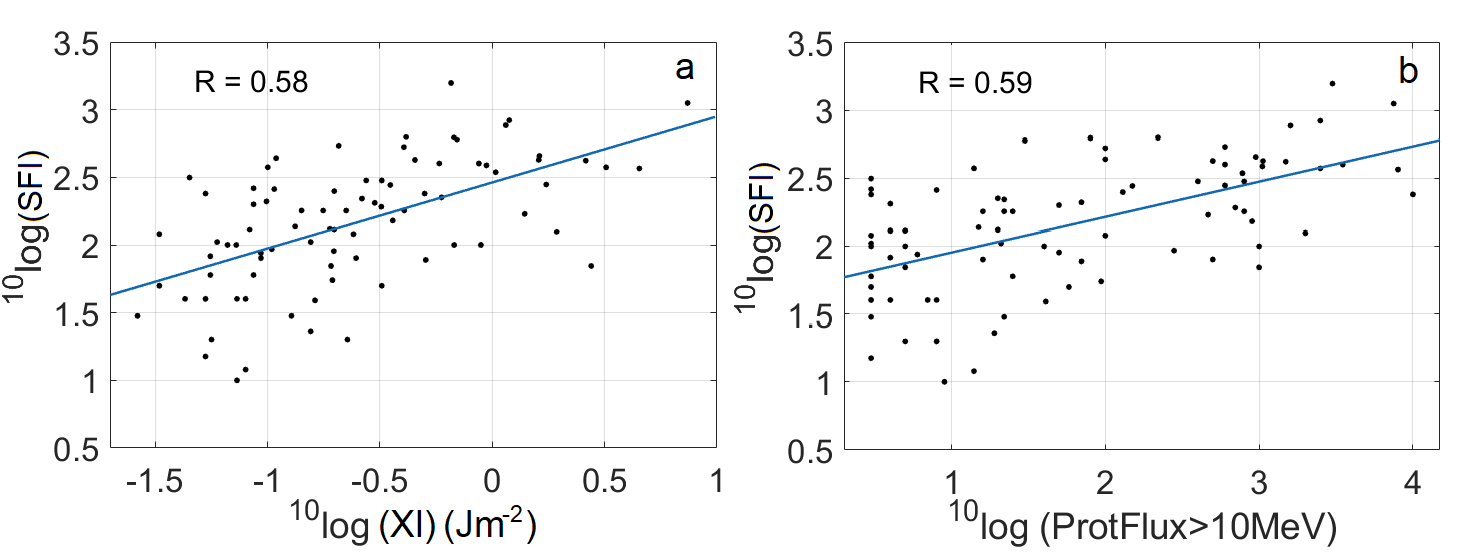}}
 \caption{a) The solar flare index (SFI) as a function of flare energy in terms of XI in logarithmic scale. b) The SFI as a function of solar proton fluxes ($>$10 MeV).}
 \label{fig:SFI_vs_X_ray}
 \end{figure}

Figure \ref{fig:SFI_vs_X_ray}a shows the solar flare index as a function of flare index XI in logarithmic scale. Here the XI is defined as
\begin{equation}
	XI = F^{max}_{0.1-0.8}\times(a+b) ,
\end{equation}
where $F^{max}_{0.1-0.8}$ is the maximum flux between 1\,--\,8 {\AA}ngstr{\"o}ms and a and b are the durations (in seconds) of the ascending phase FWQM (full width quarter maximum) and descending phase FWQM, respectively \citep{Bruevich_2020}. (The 83 flares shown in figures \ref{fig:SFI_vs_X_ray}a and \ref{fig:SFI_vs_X_ray}b are from the Solar Cycles 23\,--\,24 listed in the Table 1 by \cite{Bruevich_2020}). Figure \ref{fig:SFI_vs_X_ray}b shows the solar flare index as a function of solar proton fluxes ($>$10 MeV) affecting the earth environment after the flares. Although the connections between XI/proton flux and flare index are not straightforward the correlation coefficients are still quite high.

\begin{figure} 
 \centerline{\includegraphics[width=1.0\textwidth,clip=]{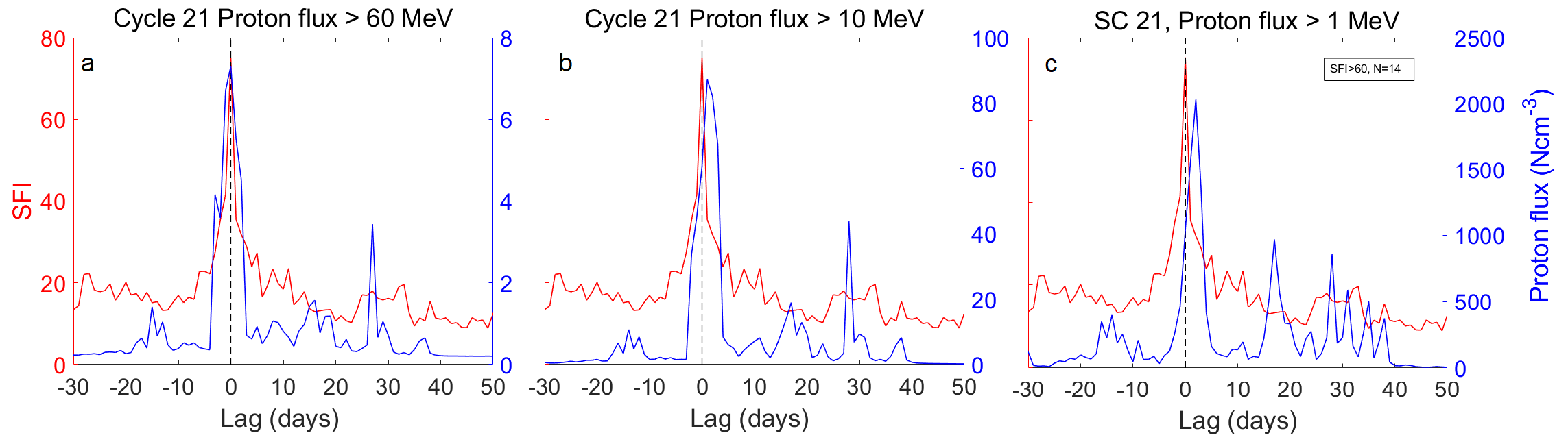}}
 \caption{The superposed epoch analyses of SFI ($>60$) and near-Earth IMF proton fluxes greater than a) 60 MeV, b) 10 MeV and c) 1 MeV for the Cycle 21.}
 \label{fig:SFI_vs_proton_flux} \end{figure}

In order to confirm the close relation between SFI and near earth proton flux, we use the superposed epoch analyses (SEA) between SFI and proton fluxes for the Solar Cycle 21, which is the strongest flare cycle having near-Earth IMF measurements. Figure \ref{fig:SFI_vs_proton_flux}a, b and c show the SEAs of SFI ($>60$) and proton fluxes greater than 60 MeV, 10 MeV and 1 MeV, respectively. There are 14 days with SFI greater than 60 in the Solar Cycle 21 (the lower level of the SFI and the number of such days during the cycle are shown in the inset of Fig. \ref{fig:SFI_vs_proton_flux}c). The zero timestamp is the day of the solar flare index exceeding 60 and we show the interval -30 to 60 days around that zero day. Note that the most energetic protons ($>$60 MeV) peak almost immediately with SFI (less than a day), while protons greater than 10 MeV peak one day after the SFI and protons greater than 1 MeV two days after the SFI. This may mean that solar flare related event, e.g. coronal mass ejection, happened already somewhat earlier than the zero moment (maximum in the SFI) in our analysis. Note also the peak after one solar rotation in all of the figures. In the proton flux (plus some smaller peaks) greater than 1 MeV, a half rotation peak also exists and is even higher than the solar rotation peak. The solar rotation peak is probably related to recurrent flare events (there is a small hump also in the SFI at one solar rotation), and the half rotation peak to less energetic protons leaving from the flare at the eastern and western limbs of the Sun.

\section{Cross-correlations of SFI and geomagnetic indices}

\begin{figure} 
 \centerline{\includegraphics[width=1.0\textwidth,clip=]{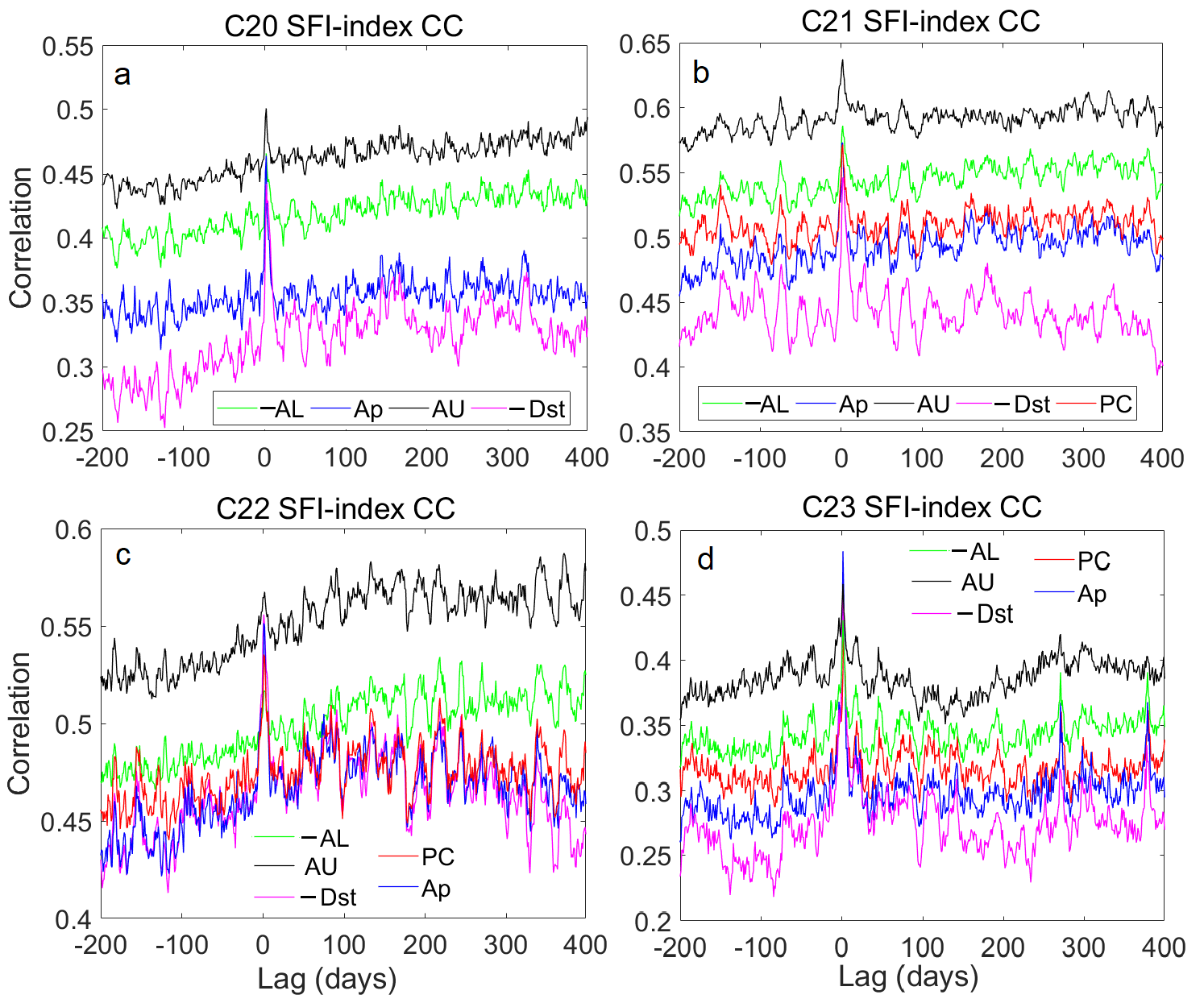}}
 \caption{The correlation coefficients between SFI and -AL, AU, PC, Ap and -Dst indices for Cycles a) 20, b) 21, c) 22 and d) 23 (PC index only for Cycles 21\,--\,23). }
 \label{fig:SFI_vs_GeomagIndex}
 \end{figure}

Earlier \cite{Takalo_2023} studied the short-lagged cross-correlation (CC) between solar flare index and geomagnetic Ap-index for Solar Cycles 19\,--\,24. Here we examine the CCs of the SFI and several geomagnetic indices using the same traditional CC-method (see definitions in \cite{Takalo_2023}). Figure \ref{fig:SFI_vs_GeomagIndex} shows these CCs for -AL, AU, PC, Ap and -Dst indices for Cycles 20\,--\,23 (PC index only for Cycles 21\,--\,23). Negative days are for each geomagnetic index preceding the SFI and positive days index succeeding the SFI. There is a peak maximizing at (positive) two to three days in all cross-correlation curves. There are, however, differences depending on the index. The AL and AU indices have only low peaks, except for the Cycle 23, which is a low activity SFI cycle. Note also that for Cycle 20 and 22 the cross-correlations between SFI and AL, AU increase still after the first peak, and especially strongly for the Cycle 22. It seems that the auroral electrojet indices do not respond as strongly to the SFI as the Ap, PC and Dst indices, although their overall CCs are higher than the CCs of the latter indices. The CCs of SFI and Ap, PC and -Dst are surprisingly similar, especially for the Cycle 22. Note also the solar rotation related 27-day peaks, which are most visible in the Cycle 21, the most active SFI Cycle of all the Cycles 20\,--\,23 \citep{Takalo_2023}. These peaks are especially clear for the Dst-index.

\section{Superposed Epoch Analyses (SEA)}

\subsection {SEAs Between SFI and Other Solar Indices}

\begin{figure} 
 \centerline{\includegraphics[width=1.0\textwidth,clip=]{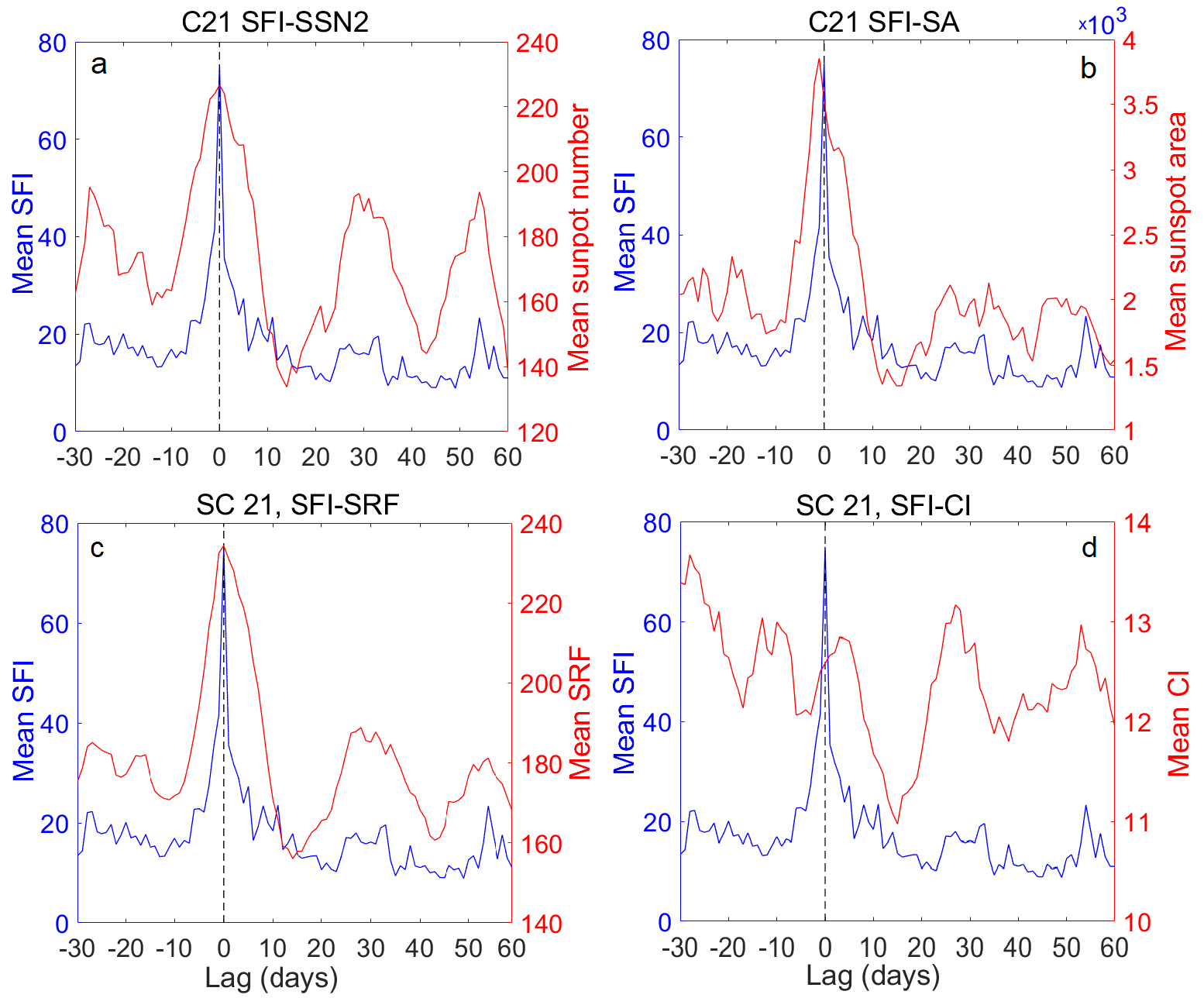}}
 \caption{The SEAs of SFI and a) sunspot number (SSN), b) sunspot area (SA), c) solar 10.7 cm radio flux (SRF) and d) coronal index of solar activity (CI) for the Solar Cycle 21.}
 \label{fig:SFI_vs_SolarIndices}
 \end{figure}

We start the investigation with superposed epoch analyses (SEA) between SFI and other solar indices. Figure \ref{fig:SFI_vs_SolarIndices}a, b, c and d show the SEA of SFI and sunspot number (SSN), sunspot area (SA), solar 10.7 cm radio flux (SRF) and coronal index of solar activity (CI) for the Solar Cycle 21, respectively. The zero time is again the day of the solar flare index exceeding 60. There are 14 events which fulfill this level in Cycle 21. It is evident that SSN, SA and SRF maximize quite simultaneously with SFI, although their peaks are broader than that of SFI. This is understandable, because flares are transient phenomena. Note that daily CI is not simultaneous with SFI and differs from the other solar indices. This is in line with the result by \cite{Takalo_2023}, who showed that the Gnevyshev gap (GG) in CI is shallower and not concurrent with the GG of other solar indices. Note also the solar rotation period especially in SSN and SRF.

\subsection {SEAs Between SFI and geomagnetic indices}

\begin{figure}
 \centerline{\includegraphics[width=0.7\textwidth,clip=]{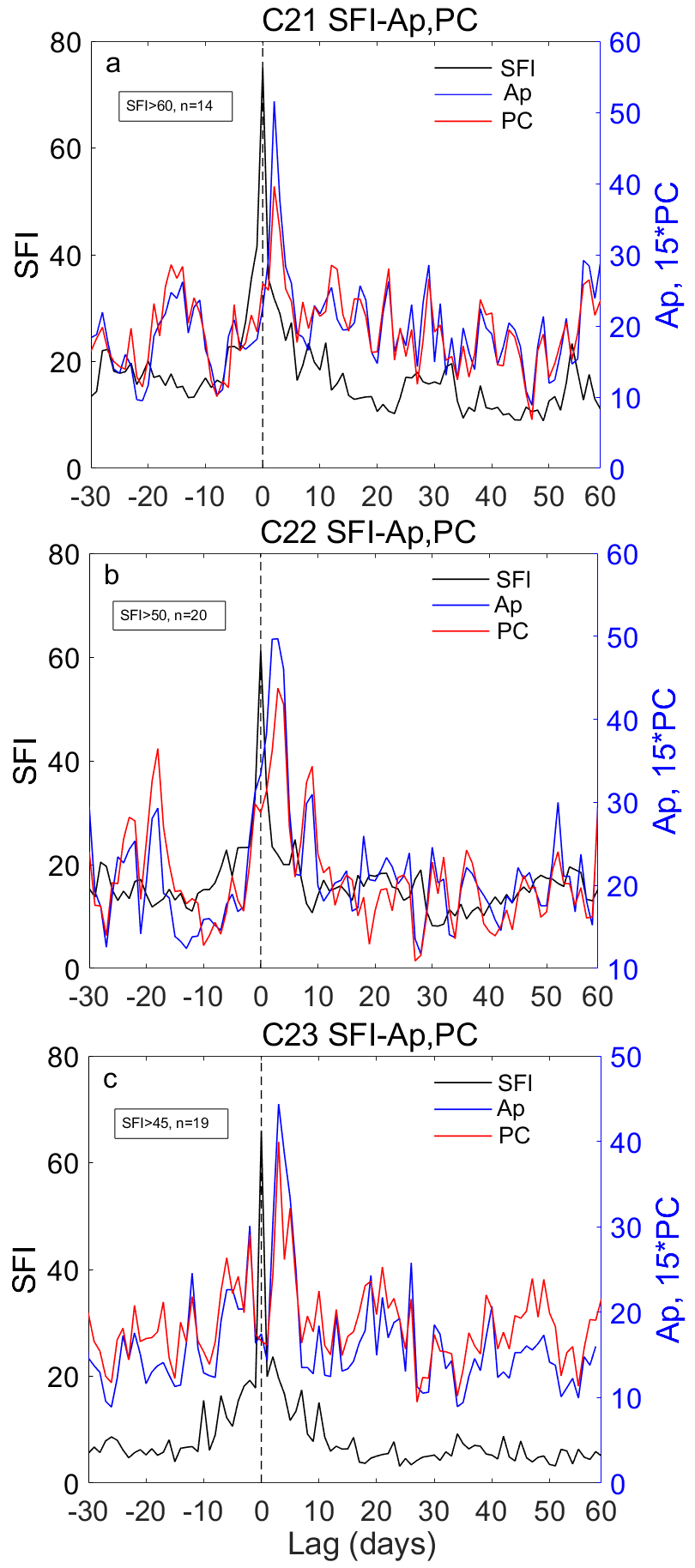}}
 \caption{The SEAs between SFI and Ap, PC-indices for Cycles a) 21, b) 22 and c) 23.}
 \label{fig:SFI_vs_Ap_PC}
 \end{figure}

\begin{figure} 
 \centerline{\includegraphics[width=1.0\textwidth,clip=]{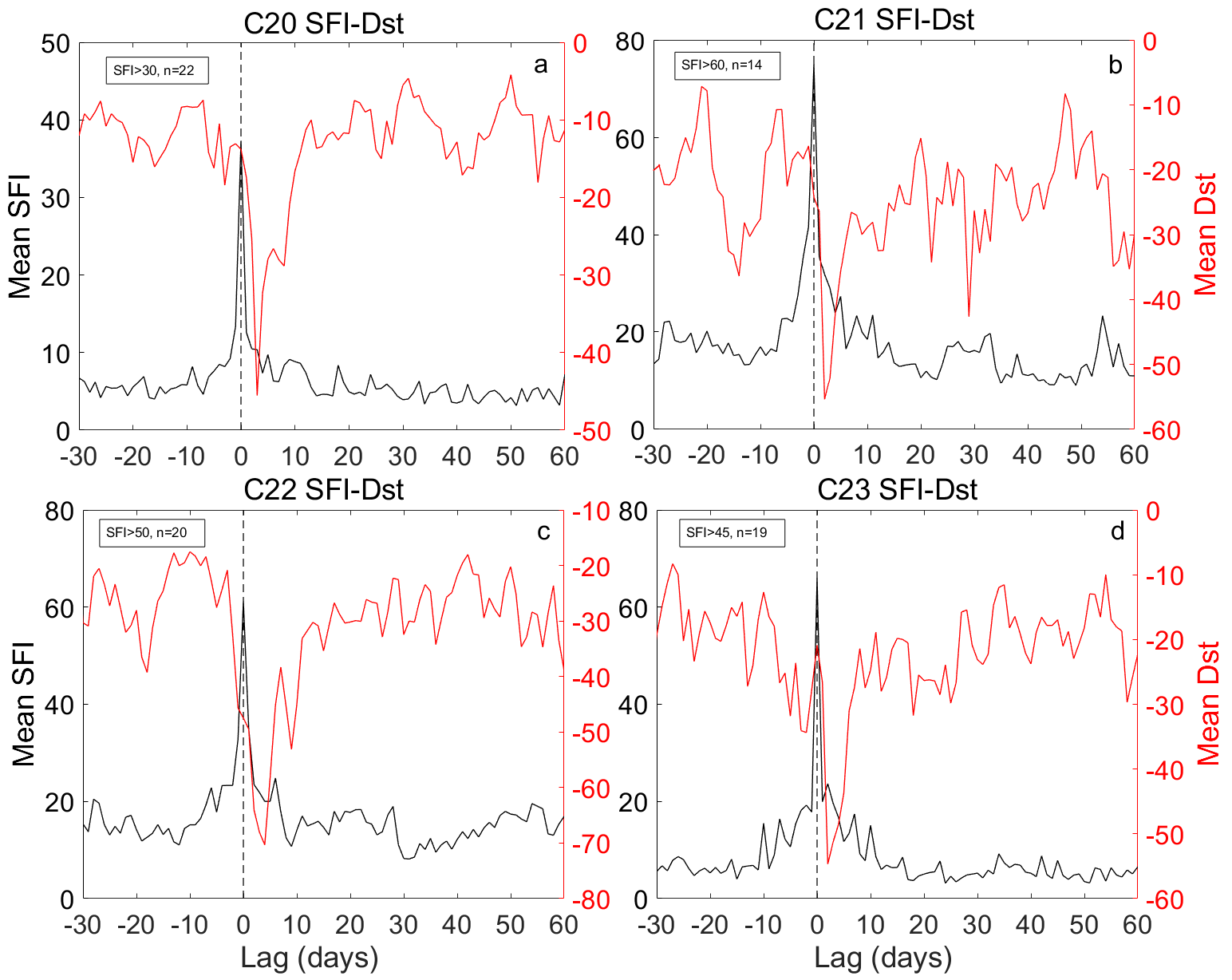}}
 \caption{The SEAs between SFI and Dst for the Cycles a) 20, b) 21, c) 22 and d) 23.}
 \label{fig:SFI_vs_Dst}
 \end{figure}

\begin{figure} 
 \centerline{\includegraphics[width=1.0\textwidth,clip=]{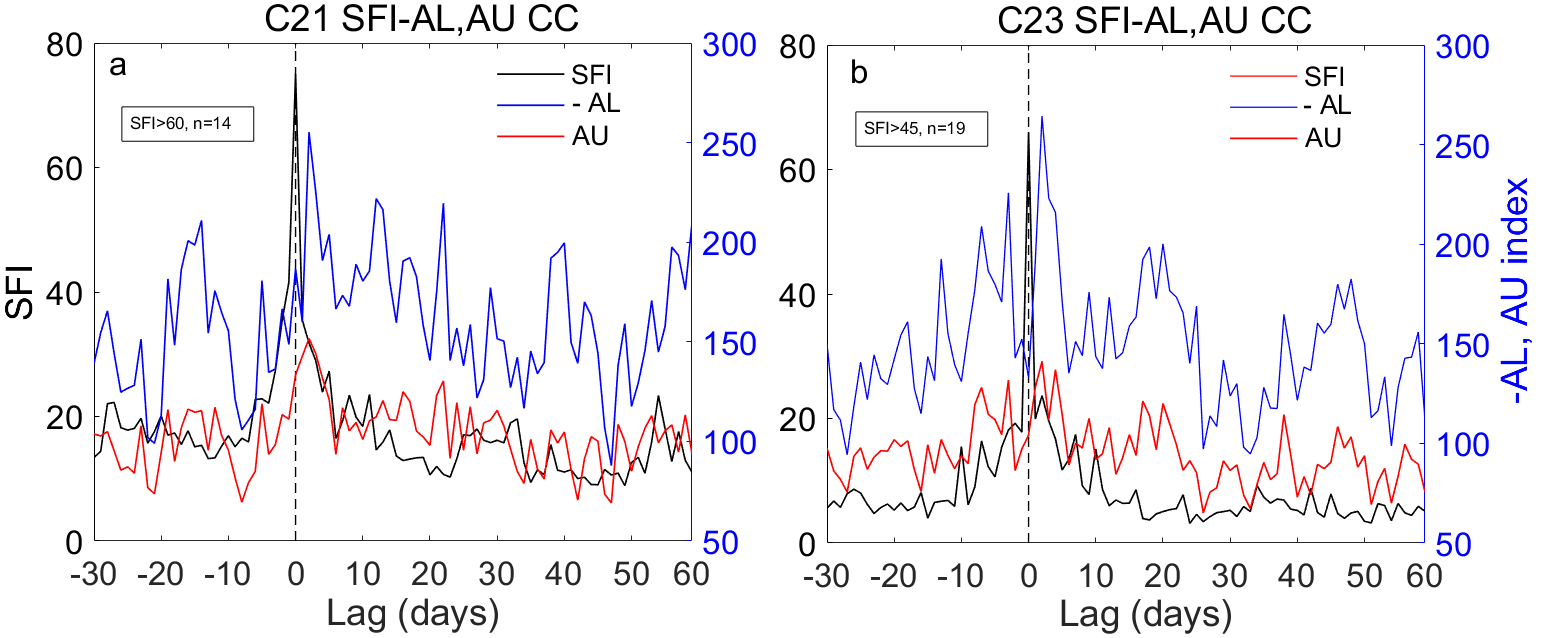}}
 \caption{The SEAs between SFI and AU, -AL indices for the Cycles a) 21 and b) 23.}
 \label{fig:SFI_vs_AU_AL}
 \end{figure}

In order to have enough superposed events between SFI and geomagnetic indices, we choose the SFI level which must be exceeded to get more than ten events for each cycle. The SFI levels (number of events) we use here are
SFI$>$60 (n=18), SFI$>$30 (n=22), SFI$>$60 (n=14), SFI$>$50 (n=20), SFI$>$45 (n=19) and SFI$>$35 (n=17) for Solar Cycles 19, 20, 21, 22, 23 and 24, respectively. 
Figure \ref{fig:SFI_vs_Ap_PC}a, b and c show the SEAs between SFI and Ap, PC-indices. (Note that PC index has been multiplied by 15 to fit to the same vertical coordinates). The lags of the Ap and PC from the zero time of SFI events is two days for the cycle 21, two to three days for the Cycle 22, and three days for the Cycle 23. Although the triggering level of SFI is not crucial for the analysis, it seems that the lag is inversely proportional to the average daily SFI strength of the Cycle, because Cycle 21 is strongest and Cycle 23 weakest SFI cycle from these three cycles \citep{Takalo_2023}.
Figure \ref{fig:SFI_vs_Dst}a, b, c and d show the SEAs between SFI and Dst-index for the Cycle 20, 21, 22 and 23, respectively. Again the, now negative, peak in Dst-index exists two days after the zero time (peak in SFI) for the Cycle 21, but after three days for the Cycle 20, which is one of the weakest SFI cycles of all measured cycles. Notice also a side peak in the Dst-index of the Cycle 20, which is minimizing eight days after zero time. This is in line with the earlier result of \cite{Takalo_2023} for the Ap-index. The Cycles 22 and 23 have now wider peaks, lasting from two to four/five days after zero time. As mentioned earlier Cycle 22 is medium strong and Cycle 23 weak SFI cycle.
Figure \ref{fig:SFI_vs_AU_AL}a and b show the SEAs between SFI and AU, AL-indices for the Cycles 21 and 23. Note that peak in -AL are separable but not as high as in the earlier indices, while AU-index does not show clear peak for Cycle 23, but a very wide shallow peak for the cycle 21. It seems that especially AU index (eastward electrojet) does not respond strongly to SFI. The peaks of the AU and AL for the even Cycles 20 and 22 are negligible and not shown here. 

\subsection {SEAs Between SFI and Cosmic Ray Intensities}

\begin{figure} 
 \centerline{\includegraphics[width=1.0\textwidth,clip=]{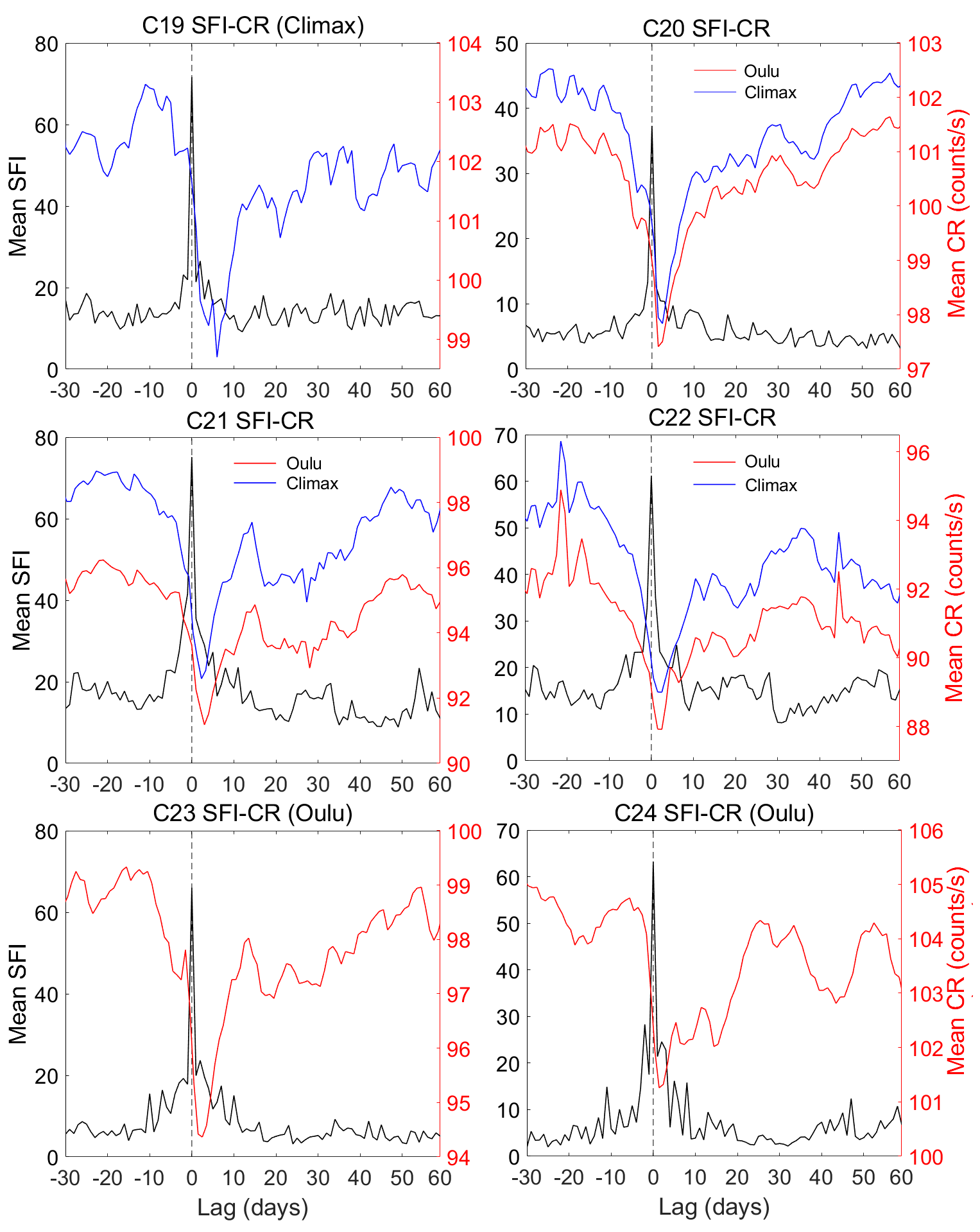}}
 \caption{The SEAs between SFI and existing CR counts of Oulu and Climax NM stations for the Cycles 19\,--\,24. }
 \label{fig:SFI_vs_CR}
 \end{figure}

\begin{figure} 
 \centerline{\includegraphics[width=1.0\textwidth,clip=]{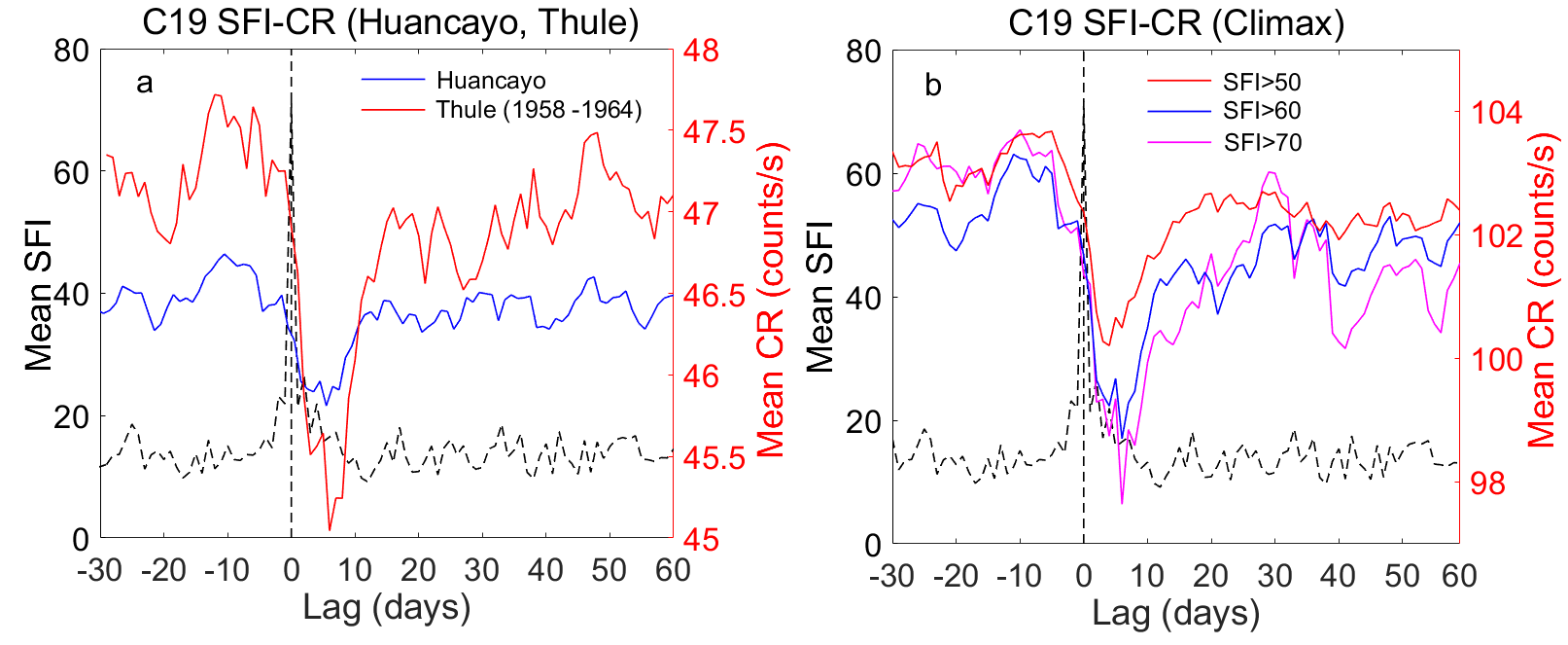}}
 \caption{a) The SEAs between SFI and CR counts of Huancayo and Thule NM stations for the Cycle 19. b) The SEAs between SFI and CR counts of Climax NM station for different lower levels of SFI during the Cycle 19.}
 \label{fig:SFI_vs_CR_Huan_Thule_Clim}
 \end{figure}

\begin{figure} 
 \centerline{\includegraphics[width=1.0\textwidth,clip=]{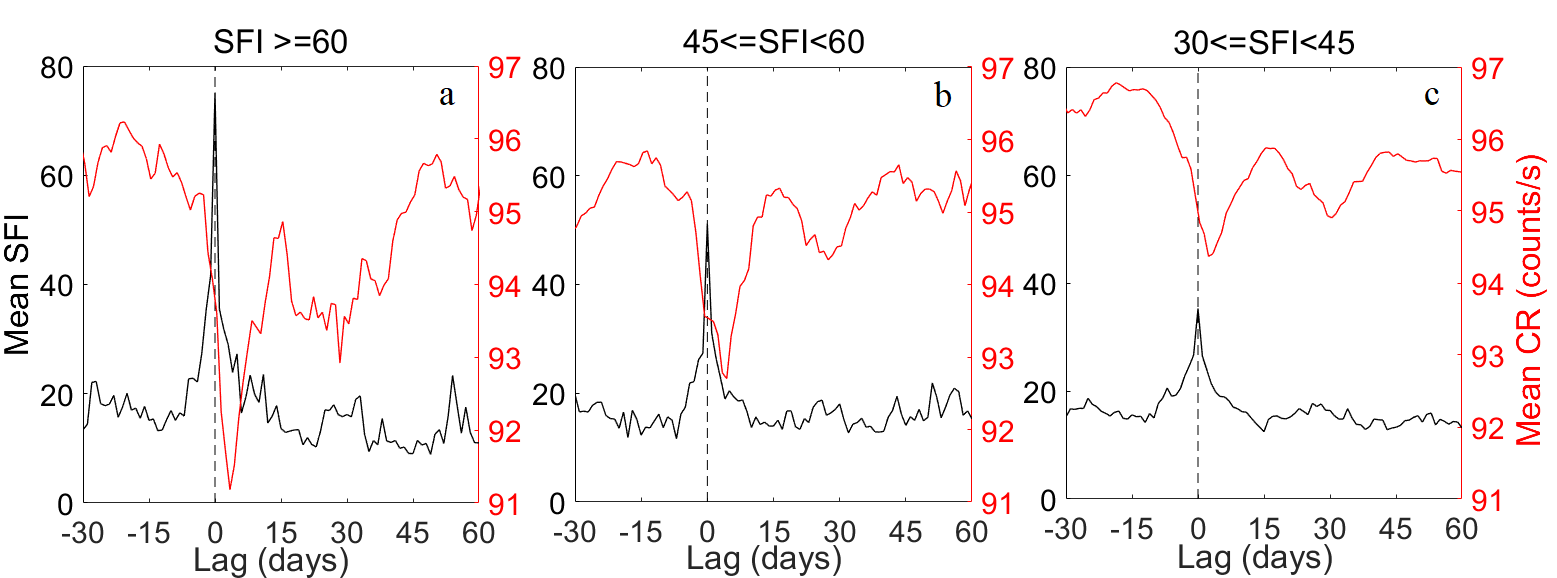}}
 \caption{The SEAs between SFI and CR counts of Oulu NM station for different intervals of SFI during the Cycle 21.}
 \label{fig:SFI_vs_CR_Oulu_diff_strengths}
 \end{figure}

Figure \ref{fig:SFI_vs_CR} shows the SEAs between SFI and cosmic ray intensity of Oulu and Climax for the Cycles 19\,--\,24. Note that Cycle 19 is caluculated only for Climax and Cycles 23 and 24 only for Oulu neutron monitor counts. The decrease in the CR intensity follows with 2-3 day lag the maximum of the SFI. It is hard to see if the lag depends of the solar cycle strength for the Cycles 20\,--\,24, and this is partly due to only one day resolution of the SFI data. On the other hand, depth of the decline seems to depend on the SFI strenght of the solar cycle,i.e., the average decline in the intensity is about 5\,\% for the Cycles 21 and 22, 4\,\% for the 20 and 23 and 3\,\% for the Cycle 24. 
The Cycle 19, however, differs from the other cycles such that it has a double-peaked decline and lasts longer than the decline of the other cycles and the deeper drop is after about 5-6 days the peak in the SFI for Climax. In Figure \ref{fig:SFI_vs_CR_Huan_Thule_Clim}a we confirm this phenomenon by plotting the Cycles 19 for Huancayo and Thule NMs. The latter is only partial cycle, because the data are available only for the years after 1958. The decline is very similar for these two NM stations than for the Climax station. The depth of the drops are, however, different, although the limit and number of SFI days is similar. The drop is only about 1.5\,\% for Huancayo while about 4.5\,\% for Thule and Climax. Note that Huancayo (not operating anymore) had a high cutoff rigidity of about 13 GeV and thus has a count rate smaller than that for Oulu, Climax and Thule neutron monitors. The high cutoff rigidity is probably the reason why the decline for the Huancayo NM count rate is small since it registers only high-energy cosmic rays, which are less modulated by the magnetospheric and the heliospheric magnetic fields. Already \cite{Forbush_1938} stated that Huancayo does not have visible seasonal variation in its early cosmic ray recordings. We, however, found a slight seasonal variation in Huancayo cosmic ray intensity for the years 1953\,--\,1984, but only -0.13\,--\,0.25\,\% from the yearly mean values. This is only one third of the seasonal variations of the Oulu and Thule cosmic ray intensities for the years 1965\,--\,1996. i.e.-0.59\,--\,0.45\,\% and -0.54\,--\,0.59\,\%, respectively. These values were calculated from 31-day smoothed average annual data.
Figure \ref{fig:SFI_vs_CR_Huan_Thule_Clim}b shows the declines for the Climax Cycle 19 for different limits of the SFI. It is evident, that the latter drop deepens when the lower limit for the solar flares is higher. This in turn shows that the first drop is due to weaker flares and the second drop to the stronger flares. This means that the lag is depending somewhat on the strength of the flares. Note that the Cycle 19 is the strongest cycle in the history of direct solar measurements.
Figure \ref{fig:SFI_vs_CR_Oulu_diff_strengths} shows the declines of the Oulu CR counts for the Cycle 21 such that the daily SFI driver is greater than 60, between 45\,--\,60 and between 30\,--\,45 in \ref{fig:SFI_vs_CR_Oulu_diff_strengths}a, b and c, respectively. It is understandable that the decline in CR is smaller when the  mean SFI is smaller. Otherwise the shape of the CR counts is very similar, but the overall level of the CR counts in the \ref{fig:SFI_vs_CR_Oulu_diff_strengths}c is higher and the peak in the deepest decrease exists one day earlier than in the others Figs. \ref{fig:SFI_vs_CR_Oulu_diff_strengths}a and b. So this shows again that the lag is depending somewhat on the strength of the flares, such that the decline is later in CR for the stronger SFI days.

\subsection {Double SEAs Between SFI and Bv$^{2}$-CR}

\begin{figure} 
 \centerline{\includegraphics[width=1.0\textwidth,clip=]{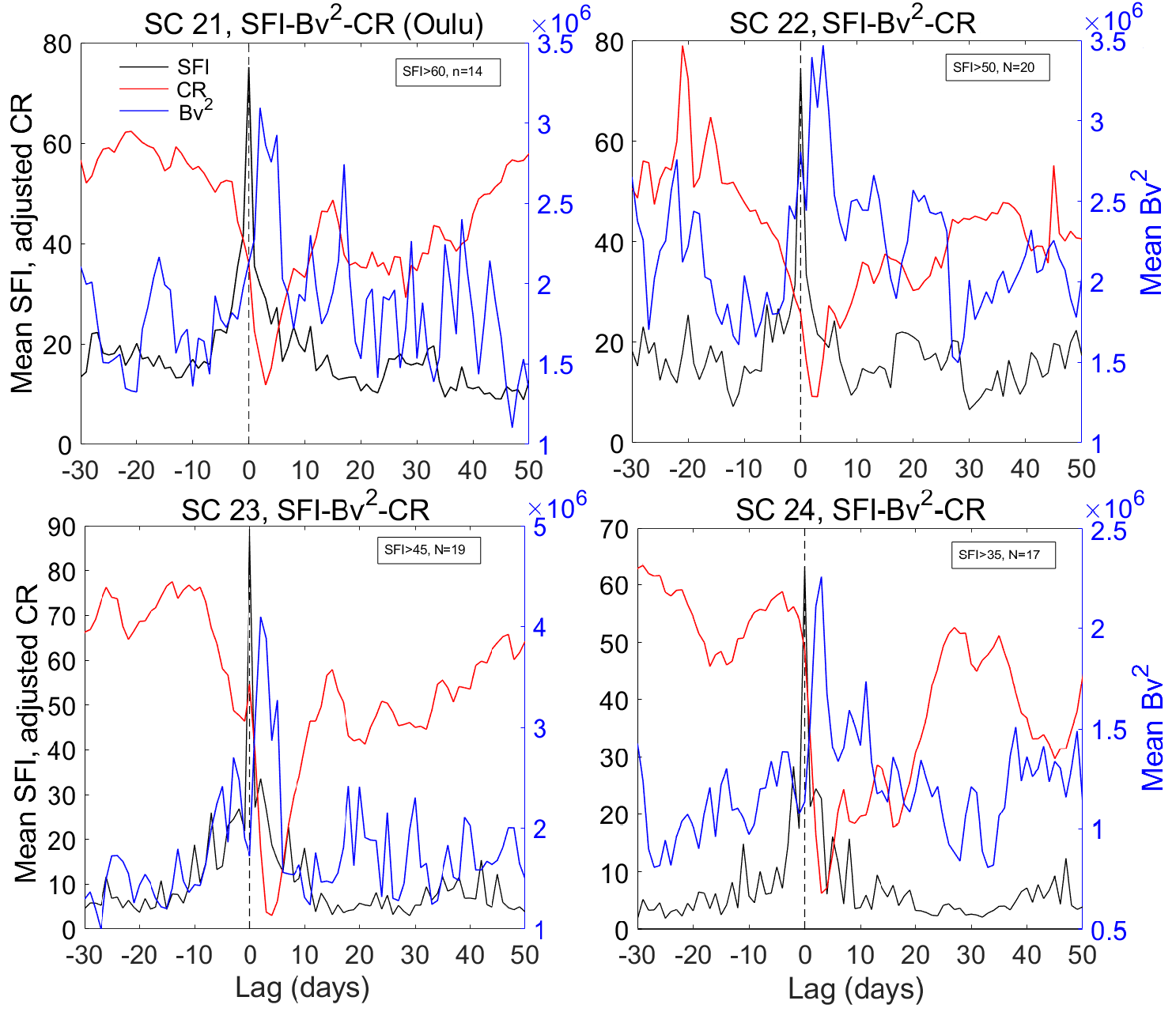}}
 \caption{The double superposed epoch analyses (DSEA) of the SFI and Bv$^{2}$-CR for the Cycles 21\,--\,24.}
 \label{fig:SFI_vs_CR_Bv2}
 \end{figure}

Figure \ref{fig:SFI_vs_CR_Bv2} shows the double superposed epoch analyses (DSEA) of the SFI and Bv$^{2}$-CR. The mean SFI is black, the mean Bv$^{2}$ blue and mean CR intensity red curve. Note that CR has been adjusted in order to fit in the same right vertical coordinates. The peaks in Bv$^{2}$ are sometimes twofold but more or less simultaneous or somewhat earlier than the decline in CR intensity. Especially for odd Cycles 21 and 23 the first peak in Bv$^{2}$ is one day earlier than the deepest drop in CR, and exists two days after SFI maximum peak.

\newpage

\section{Conclusions}

We have shown that the most energetic protons ($>$60 MeV) in near-Earth IMF have a peak almost 
immediately (less than a day) after the peak in solar flare index (SFI) , while protons greater than 
10 MeV peak one day after the SFI and protons greater than 1 MeV two days after the SFI. Furthermore, 
we have shown that the SFI is related to the proton flux ($>$10 MeV) with correlation coefficient 0.59.

According to our SEA, the sunspot number, sunspot area and solar 10.7 cm radio flux and coronal are very well synchronized with solar flares, although their peaks are understandably broader than that of the SFI. On the other hand the coronal index of solar activity (CI) is not simultaneous with the SFI.

We show that the geomagnetic indices AU, -AL, PC, Ap and -Dst have positive peaks after two to three 
days in all cross-correlation curves. The -AL and AU indices have, however, only low peaks,
except for the Cycle 23, which is a low activity SFI cycle. It seems that the auroral
electrojet indices do not respond as strongly to the SFI as the Ap, PC and Dst
indices, while interestingly their overall CCs are higher than the CCs of the latter indices. This is especially true for the AU, i.e eastward electrojet.

The aforementioned results are confirmed using superposed epoch analysis (SEA) for the SFI and geomagnetic indices.
In this analysis we see again that the electrojet indices do not respond as strongly to the SFI driver than
the other geomagnetic indices. Actually, the response of the eastward electrojet, AU, is negligible in this analysis.

The SEAs of the SFI and cosmic ray counts (CR) show that the decline in the
CR intensity follows also with 2-3 day lag the maximum of the SFI for the Solar Cycles 20\,--\,24. 
The decline seems to depend on the SFI strenght of the solar cycle, i.e., the average
decline in the intensity is about 5\,\% for the Cycles 21 and 22, 4\,\% for the 20 and
23 and 3\,\% for the Cycle 24. The Cycle 19, however, differs from the other cycles
such that it has double-peaked decline and lasts longer than the decline of the other
cycles. The deeper drop, which is due to stronger SFIs, exists after about 5-6 days the peak in the SFI. 
This phenomenon is similar for the SEAs of the Cycle 19 for Climax, Huancayo and Thule stations. We suppose that the strength of the solar flares affects to the heliospheric modulation of the cosmic rays such that the lag is somewhat longer for more intense flares.

The double superposed epoch analyses show that the response of IMF Bv$^2$ and CR to SFI are quite simultaneous or sometimes Bv$^2$ may peak somewhat earlier than the decline in CR. This is, however, not so clear while Bv$^2$ is sometimes double peaked. The first peak in Bv$^2$ exists anyhow about two days after the peak maximum in the SFI. These results are also somewhat vague, because the resolution of the data are only one day.

From the aforementioned analyses it is clear that, when we observe high activity in white light H$\alpha$ flares, there is a possibility
that magnetic field of the Earth experiences soon strong disturbances. We, however, have shown that there are couple of days to react to this situation.

\begin{figure} 
 \centerline{\includegraphics[width=1.0\textwidth,clip=]{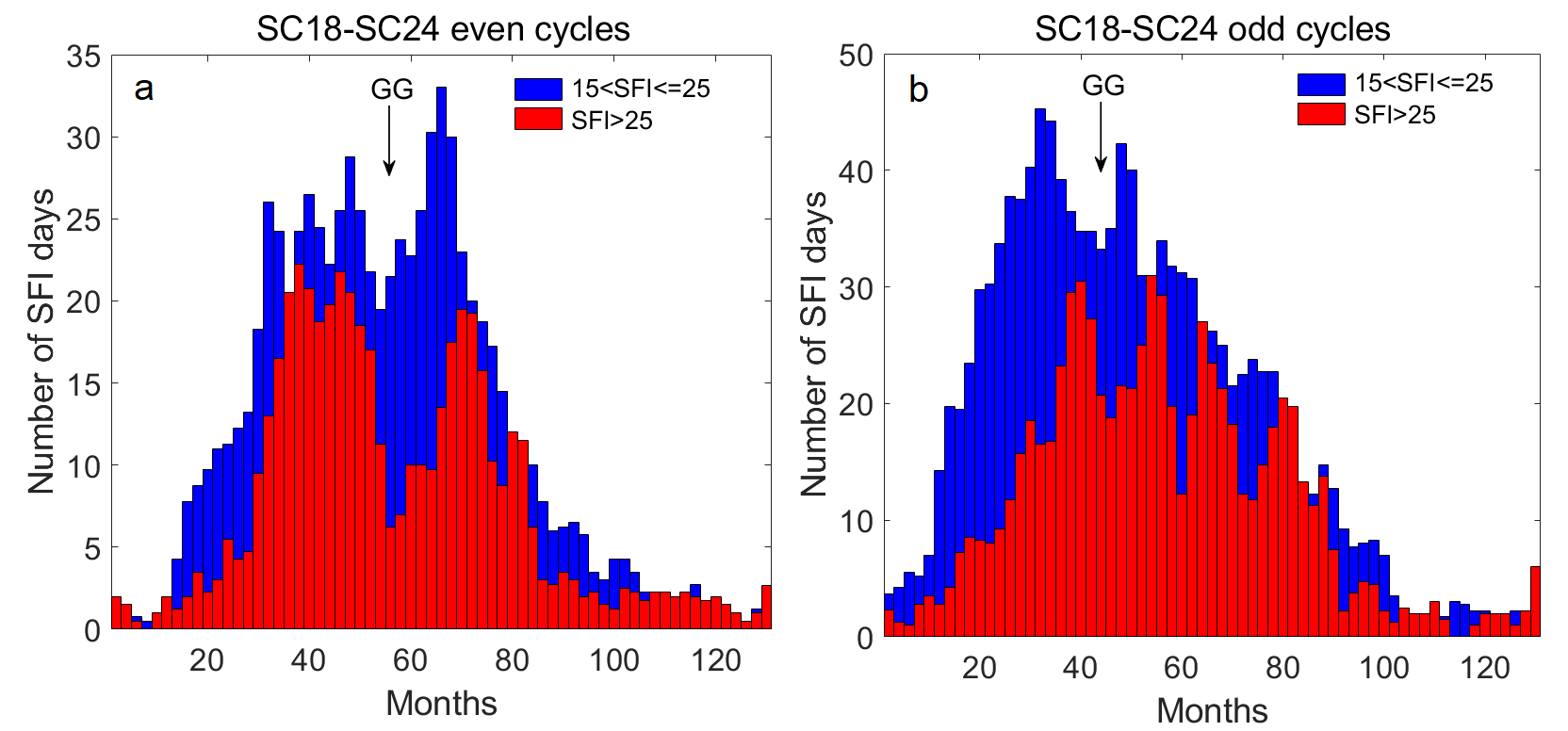}}
 \caption{The histograms of amounts for strong and very strong SFI days a) for the even Cycles 18\,--\,24 b) for the odd Cycles 18\,--\,24. GG shows the Gnevyshev gap.}
 \label{fig:Histograms}
 \end{figure}

Furthermore, \cite{Takalo_2023} has shown that there is a huge Gnevyshev gap (GG) in the SFI of the average
solar cycle, especially for the even cycles. Figure \ref{fig:Histograms}a and b show the histograms of
strong and very strong SFI days for the even and odd Solar Cycles 18\,--\,24, respectively. The cycles here have been resampled such that each cycle has 130 months. In the histograms each bar represents two months and 
strong days are shown as blue bars (15$<$SFI$\leq$25) and very strong days as red bars (SFI$>$25).
It is evident that the Gnevyshev gap is a very strong phenomenon in the SFI of the even cycles, while less strong 
in the odd cycles. In addition, the GG occurs somewhat earlier in the odd cycles. We know also that GG exists also in the geomagnetic indices and is again more visible in the even cycles than in the
odd cycles \citep{Takalo_2021_1}.

The GG in the activity of the Sun is so clear that it must have
influence on the space-weather as suggested also earlier by Storini et al. (2003). If we
can predict the forthcoming length of the solar cycle, it is possible to presume
the GG related less active window in the IMF and solar-terrestrial interaction. Furthermore, we can utilize this 
less vulnerable window in space missions. In fact, according to our calculations, the preparatory Apollo flights 8\,--\,10 were done during the GG period of the Cycle 20, although it was not known during the missions. We were also lucky that the rest of the Apollo missions were launched in the descending phase of one of the weakest flare cycles, i.e. Cycle 20 \citep{Takalo_2023}.

%

%

%
\newpage

 \begin{acks}
I acknowledge the NMDB database (www.nmdb.eu), founded under the European Union’s FP7 programme (contract no. 213007), and the teams of the Climax, Huancayo, Oulu and Thule neutron monitors for providing data. The Solar Flare index data were obtained from https://dataverse.harvard.edu/dataset.xhtml?persistentId=doi:10.7910/ \newline DVN/U5GR3D. Part of the Flare Index Data used were calculated by T.Atac and A.Ozguc from Bogazici University Kandilli Observatory, Istanbul, Turkey. The dates of cycle minima were obtained from from the National Geophysical Data Center, Boulder, Colorado, USA (https://www.ngdc.noaa.gov/stp/space-weather/solar-data/solar-indices/sunspot-numbers/ \newline cycle-data/table\_cycle-dates\_maximum-minimum.txt). The newly reconstructed corona indices were obtained from www.ngdc.noaa.gov/stp/solar/corona.html. The corona index for Solar Cycle 24 was downloaded from www.suh.sk/obs/vysl/MCI.htm. The SSN2 data has been downloaded from www.sidc.be/silso/datafiles, and the solar radio flux data from lasp.colorado.edu/ \newline lisird/data/penticton\_radio\_flux/. Definitive values of Ap are provided by GeoForschungs Zentrum (GFZ) Potsdam. The OMNI2 data are downloaded from https://spdf.gsfc.nasa.gov/ \newline pub/data/omni/low\_res\_omni/. I am also grateful to I. Usoskin for useful communication.
 \end{acks}

%
%
%
%
%
%
%

%
%
\bibliographystyle{spr-mp-sola}
\bibliography{references_JT_SolPhys}  
%
%
%
%

\end{article} 
\end{document}